\documentclass[preprint,showpacs,showkeys,preprintnumbers,amsmath,amssymb]{revtex4}

\usepackage{graphicx}
\usepackage{dcolumn}
\usepackage{bm}

\begin{document}

\preprint{APS/123-QED}

\title{Monte Carlo simulation for statistical mechanics model of ion channel cooperativity in cell membranes}

\author{Riza Erdem$^1$}
\author{Ekrem Aydiner$^2$}%
\email{ekrem.aydiner@deu.edu.tr}
\affiliation{%
$^1$Department of Physics, Gaziosmanpa\d{s}a University, Tokat 60250, Turkey \\
$^2$Department of Physics, Dokuz Eylul University, Izmir  35100,
Turkey
}%

\date{\today}

\begin{abstract}
Voltage-gated ion channels are key molecules for the generation and
propagation of electrical signals in excitable cell membranes. The
voltage-dependent switching of these channels between conducting and
nonconducting states is a major factor in controlling the
transmembrane voltage. In this study, a statistical mechanics model
of these molecules has been discussed on the basis of a
two-dimensional spin model. A new Hamiltonian and a new Monte Carlo
simulation algorithm are introduced to simulate such a model. It was
shown that the results well match the experimental data obtained
from batrachotoxin-modified sodium channels in the squid giant axon
using the cut-open axon technique.
\end{abstract}

\maketitle

\section{\label{sec:level1}Introduction}
Ion channels are a class of proteins that reside in the membranes of
all biological cells and form conduction pores that regulate the
transport of ions into and out of cells  \cite{Hille}. Not only are
ion channels important in their biological purpose, they are also
extremely interesting for their possible use in bio-electronics,
more specifically as a component in a new class of bio-sensors.
Different channels are classified by their gating mechanisms. Some
are voltage-gated; others are ligand-gated (e.g., neurotransmitter
gated channels) and yet other channels can be gated by heat or even
mechanical stress. These gating mechanisms and different forms of
cooperativity between ion channels are central issues in biological
physics because of the dominant role channels played in the control
of key cellular processes \cite{Manivannan,Keleshian,Ghosh}.
Cooperative behavior in ion channels arises from direct, energetic
interactions between components of a system. Such interactions give
rise to a sigmoidal dependence between the state of the system
(e.g., ligand binding to a receptor) and the thermodynamic variables
(e.g., ligand concentration"). Two levels of interaction can take
place in a system of ion channels: local cooperation between
subunits of a macromolecular channel protein, and non-local
cooperation between different channel proteins.

Many studies have considered the cooperative interactions between
ligand-gated ion channels. Such interactions give rise to a
sigmoidal dependence between the ligand binding to a receptor and
ligand concentration. Ligand-gated channel current is usually
derived in equilibrium and thermotropic type models such as the Hill
\cite{Hill} and Langmuir \cite{Langmuir} equations. Also, the Ising
model of statistical mechanics and lattice type theories employed in
this field are of equilibrium and thermotropic with a chemical
potential modified to accommodate for the dependence on ligand
concentration \cite{Liu}. In these works, local cooperativity was
illustrated by the Hill equation and the contribution from
short-range cooperativity to the Hill coefficient has also been
evaluated in a nearest-neighbor interaction, Ising type model.
Contrary to the Hill, Langmuir, Ising, and other stationary type
models, lyotropic, non-stationary, non-equilibrium, long-range
interaction models were derived, recently \cite{Matsson1,Matsson2}.
These models were based on the DNA replication and the cell cycle
\cite{Matsson3} and a comparison of their model results with an
Ising-like model provided a methodology to obtain the
non-equilibrium scaling dependence of Ising-like  models on the
reactant concentration. Besides these works, Guo and Levine
\cite{Guo} introduced a phenomenological model for the clustering
which is due to an interaction between nearest-neighbor receptors.
They described the clustering by the statistical mechanics of a
simple lattice Hamiltonian and calculated a phase diagram. Zafar et
al. \cite{Zafar} presented a statistical mechanics model for the
interaction between the neurotransmitter and receptor ion channels.
Based on the model, an equation for concentration response curves
was derived and the model provided good fits to measured curves.

Voltage-sensitive membrane channels show also cooperative behavior
as evident from electrophysiological experiments
\cite{Iwasa,Vijayvergiya}. Using the principles of statistical
mechanics, different approaches were proposed to account for the
inter-channel cooperativity. Ghosh and Mukherjee \cite{GhoshS1} have
suggested a statistical mechanical approach for a microscopic
analysis of these systems. They developed a model of the Zimm-Bragg
type for a membrane with a large number of channels behaving
cooperatively and tested their model with voltage-dependent
conductance data for gap-junction channels in embryonic cells. Ghosh
\cite{GhoshS2} extended their analysis for the relaxation of
channels to deal with the time-dependent observations. Yang et al.
\cite{Yang} proposed a statistical mechanical model of cell membrane
potassium and sodium ion channels in squid giant axon. The model
incorporated interactions between the tissue electric field and the
respective ion channels and was equivalent to the familiar lattice
gas model in statistical mechanics. Under a mean-field
approximation, the maximum fractions of potassium and sodium
channels were obtained by solving a self-consistent nonlinear
equation. The model produced an excellent fit to experimental
observations for maximum fractions of potassium and sodium
conduction under a static external stimulus. Erdem and Ekiz
\cite{Erdem1} developed the Bethe lattice version of this model
based on the recursion technique used for magnetic systems and
showed the collective effects introduced by trans-membrane voltage.
Recently, Erdem \cite{Erdem2} used the one-dimensional Ising model
formalism to study the collective equilibrium behavior of
voltage-gated channels. In Ref. \cite{Erdem2}, effects of negative
and positive cooperativity on the open probability of a
batrachotoxin-modified (BTX) sodium channels were discussed. While
there is vast literature devoted to statistical mechanics modelling
of ion channel cooperativity, these have not been adequately
addressed using computational methods \cite{Guo}.

Computational studies can make meaningful contributions to our
understanding of biological ion channels \cite{Roux}. A wide variety
of computational methods, such as molecular dynamics simulations
\cite{Smith}, kinetic rate models \cite{Schumaker}, continuum
electrostatic Poisson-Boltzmann theory \cite{Ranatunga} and Brownian
dynamics \cite{Millar}, provided a virtual route for interpreting
experimental observations on relating a channel structure to its
function. Besides these works, the Monte Carlo (MC) simulation
technique was used to model the noise and stochastic resonance in
voltage-gated ion channels \cite{Adair}. Adair \cite{Adair} took the
Hodgkin-Huxley (HH) description of sodium and potassium channels in
the squid giant axon as a basis of calculation and used relations
from statistical mechanics when the channel states were separated by
a Gibbs energy. In the HH formulation of the channel dynamics, the
state of the domains depends only on the potential difference and
not on the neighboring channel states \cite{Hodgkin}. Im et al.
\cite{Im} combined the Grand Canonical MC simulations with the
Brownian dynamics to simulate movement of ions in membrane channels.
This approach provided a framework for simulating ion permeation in
context of detailed microscopic models. Similarly, an algorithm in
which kinetic lattice Grand Canonical MC simulations were combined
with mean-field theory was presented to calculate ion currents in
model ion channel system by Hwang et al. \cite{Hwang}.

On the other hand, the long-range coupling between ion channels in
squid giant axons originates from specific distribution of sodium
channels in the membrane \cite{Hanyu} and also from global
capacitance coupling \cite{Fox}. Transmembrane voltage is a global
membrane property determined in part by the membrane capacitance.
Through their transition rates, it globally couples the independent
ion channels \cite{Fox}. When the membrane channel number is small,
the individual channels open and close independently, producing a
very noisy transmembrane potential. As the membrane channel number
increases, a transition to regular, collective behavior occurs. The
regular behavior in globally coupled stochastic elements has been
experimentally observed and analyzed by Sherman et al.
\cite{Sherman}. The molecular origin of the spatial long-range
interactions in squid giant axons arises from the specific
distribution of sodium channels \cite{Hanyu} and the BTX is a
neurotoxin that plays an important role in understanding the
molecular properties of sodium channels in excitable membranes.
Among various toxins studied the BTX is one of the most potent and
specific activators of sodium channels, causing them to remain open
at the resting membrane potentials. The effects of BTX on sodium
channels in hybrid neuroblastoma cells were studied by using voltage
clamp method by Huang et al. \cite{Huang}. BTX-modified sodium
channels activated with first-order kinetics and, over most of the
potential range, activated more slowly. Iwasa et al. \cite{Iwasa}
analyzed the current records from voltage-clamped membrane patches
containing two BTX-modified sodium channels in hybrid neuroblastoma
cells to determine whether these channels are identical and
independent. Their results showed a clear discrepancy with binomial
distribution and suggested that the explanation of this discrepancy
was negative cooperativity between two channels in a patch. The
negative cooperativity described in Ref. \cite{Iwasa} was based on a
slower rate of channel opening when the neighboring channel was open
than it is closed and this was considered as a useful property to
relate channel structure to channel function in future attempts.
These cooperative interactions do not occur between every pair of
channels and it is difficult to imagine interactions between
channels separated by large distances. Of nine two-channel patches
observed in Ref. \cite{Iwasa} five patches contained clearly
interacting channels, two patches contained two independent,
identical channels, and the two patches showed a behavior consistent
with either non-independence or non-identity. The interacting
channels were the majority when a large number of two-channel
patches were examined. The analysis of two-channel patches showed
relatively simple kinetics involving two closed states and one open
state.

Indeed, the statistical mechanics model of voltage-sensing ion
channel cooperativity in cell membranes has significant role in
theoretical studies since it makes meaningful contributions to our
understanding behavior of biological ion channels. Therefore, in
this paper, in order to obtain open probability of ion channels at
different voltages and temperatures, we study the simulation model
of the gating kinetics of single squid axon BTX-modified Na+
channels \cite{Correa} using MC simulation technique. The
BTX-modified sodium channels in the squid giant axon have simple
kinetics involving two states as open ($O$) or close ($C$) and their
steady state activation contains our gating parameters (e.g., gating
charge per channel and the midpoint values for the membrane
potential), hence they are convenient for our simulation purpose.

The paper is organized as follows: In Sect.\,2. We present the model
with a Hamiltonian and suggest the simulation method to calculate
the open probability of the BTX-modified sodium channels.
Comparisons between MC simulation and experimental results obtained
using cut-open axon technique and the effects of cooperative
interactions on the simulation results are given in Sect.\,3.
Finally in Sect.\,4. Our conclusions are summarized and discussed.

\section{The Model and Simulation Method}

We mentioned above that ion channels which have different biological
and physico-chemical structures with different gating mechanisms are
located on the membrane surface as grown into the cell. These
channels can be either open ($O$) or closed ($C$). There is some
structure or the property of the channel that is concerned with the
transition between these two states, and the word \emph{gate} is
used to describe this concept. When the gate is open the ions flow
through the channel, and when it is closed they can not. Gating is
the process whereby the gate is opened and closed. There may be a
number of different closed and open states, so the gating processes
may involve a number of different sequential or alternative
transitions from one state of the channel to another. On the other
hand, \emph{modulation} occurs when some substance or agent affects
the gating of the channel in some way.

The behavior of a simple two-state channel is described by a simple
two-state system in which each channel is opened by the movement of
a single gating particle which carries a charge. At any moment the
particle is in one of two positions, 1 and 2, and these are
associated with closed and open states, respectively. The positions
$1$ and $2$ correspond to two wells in an energy profile, and there
is a single energy barrier between them. In this theoretical
framework, in a population of $N$ identical channels in the
membrane, the number of channels in the closed and open states is
indicated by $n_{1}$ and $n_{2}$, respectively. $n_{1}$ is the
occupation number of gating particles in position $1$ with energy
$\varepsilon_{1}$ and $n_{2}$ is the occupation number of gating
particles in position $2$ with energy $\varepsilon_{2}$ so that
$n_{1}+ n_{2}=N$. A simple expression for the internal energy of
such a system in the presence of a membrane potential is given by
\cite{Aidley}
\begin{equation}
E=\sum_{i=1}^{2}n_{i}\varepsilon_{i}+ze_{0}n_{1}V
\end{equation}
where $z$ is the number of charges on gating particle, $e_{0}$ is
the elementary electronic charge and $V$ is the potential
difference, also called the membrane potential. This expression of
internal energy does not include interactions between ion channels.
However, some theoretical models have supposed by many authors as
including interactions between ion channels. At this point, we must
remark that one of the most fascinating models is a statistical
model based on spin systems \cite{Yang}. By analogy between channel
and spin systems it was considered \cite{Yang,Erdem1,Erdem2} that
membrane is a two-dimensional sheet and assume that the channels are
located on a finite square lattice in the membrane with each channel
having two states (open and closed) and incorporating the
interactions between these states. In this context, the energy of
the channel system was defined in the form
\begin{equation}
E=-J\sum_{\left\langle ij\right\rangle }^{N}
\sigma_{i}\sigma_{j}-ze_{0}\left(
V-V_{0}\right)\sum_{i=1}%
^{N}\sigma_{i}
\end{equation}
where $\sigma_{i}$ can take the values $0$ and $1$ corresponding to
$C$ and $O$ states, respectively, and $N$ denotes total number of
channels on the membrane surface. Also $J$ represents the
interaction energy between a pair of nearest neighbor channels,
denoted by ${\langle ij \rangle }$, and $V_{0}$ [equals to $-\left(
\varepsilon_{1}-\varepsilon_{2}\right) /ze_{0}$] is the voltage at
which half of the channels are open. In type of this study, open
probability for interactive system has been obtained using mean
field approximation. In fact, it is shown that the mean field
solutions of theoretical models \cite{Yang,Erdem1,Erdem2} with
nearest-neighbor interacting states give compatible results with the
experimental studies.

Indeed, a channel system can be represented by a spin system since a
membrane is essentially a surface phenomenon and a channel has two
possible states, and representing the distribution of channels over
the cellular membrane on a two-dimensional array with each channel
having two states and with interactions we are led to a generalized
two-dimensional spin model. However, we have seen in this study that
Eq.\,(2) is not a simulation model of channel systems. Meaningful
results of Eq.\,(2) in the mean field approximation is only obtained
if spin variable $\sigma_{i}$ takes on the values of $0$ and $+1$
instead of $-1$ and $+1$. But simulation of Eq.\,(2) with spin
$\sigma_{i}=\pm1$ produces magnetization curve which changes in the
interval [$-1$,$+1$], and magnetization of a spin system does not
corresponds to open probability for a channel system. Therefore, in
order to obtain open probability of a channel system, we here
introduce a new model based on spin system. Note that, in this
model, we also visualize the membrane as a two-dimensional sheet,
and we assume that the channels are located on a finite square
lattice in the membrane with each channel having two states (open
and closed) and incorporating the interactions between these states
like in mean field model, and we assume that, in a spin model with
$\sigma_{i}=\pm1$, the probability of inversion of a spin from $-1$
to $+1$ corresponds to the open probability $P_{o}$ of a channel on
the cell surface. Based on this assumption, we can simulate the spin
system to obtain open probability of channels on the cell membrane.
To simulate the open probability of such a channel system, a new
Hamiltonian was introduced as
\begin{equation}
E=J\sum_{i=1}^{N}\sigma_{i}P_{o}+ze_{0}\left(  V-V_{0}\right)  \sum_{i=1}%
^{N}\sigma_{i}
\end{equation}
where $P_{o}$ is defined in terms of magnetization of spin system as
\begin{equation}
P_{o}=\left(  \frac{1}{2}+\frac{\widetilde{m}}{2}\right)
\end{equation}
where $\widetilde{m}$ is the average magnetization per spin, and it
is given by
\begin{equation}
\widetilde{m}=\frac{1}{N}\sum_{j}^{N}\sigma_{j} \ .
\end{equation}
In these equations $P_{o}$ represents open probability of channels,
$\sigma_{i}$ (and $\sigma_{j}$) can take the values $+1$ and $-1$
corresponding to $O$ and $C$ states, $N$ denotes total number of
spins on the lattice which corresponds to the number of channels on
the membrane surface. Energy of a given channel system in this model
in terms of overall probability i.e., open probability $P_{o}$ which
is a global variable. Similarly, the energy of a single channel can
be written depends on overall probability as
\begin{equation}
E_{i}=J\sigma_{i}P_{o}+ze_{0}\left(  V-V_{0}\right)  \sigma_{i} \ .
\end{equation}
Since the interaction is of infinite-range in Eq.\,(3) a global
coupling between channels are introduced for the system. In the case
$J=0.0$, the first term is absent in Eq.\,(3) this means that the
global coupling disappears, hence open probability is calculated in
terms of the second term. However, for $J \ne 0$ values, global
coupling in the first term shows effects on the open probability. It
will be shown in the next section that the energy proposed above,
namely Eq.\,(3), and its single channel modification Eq.\,(6) is a
correct approximation for calculating the probability $P_{o}$.

We deal with the probability of flipping of down spins which
corresponds to the open probability of close channels in this model.
Therefore, during the simulation we focus only on the probability of
flip instead of numerical difference between up and down spins,
i.e., net magnetization. Here we will simulate the model described
in Eq.\,(3) using a new MC algorithm to calculate the open
probability $P_{o}$ since direct simulation of Eq.\,(2) with
standard MC method is not suitable. Our simulation algorithm shows a
few significant differences from standard Metropolis MC Algorithm.
Therefore, before we give our simulation algorithm steps, we remind
briefly standard Metropolis MC algorithm steps.

It is known that, in the standard Metropolis MC algorithm, the
Boltzmann probability function $p$ is defined as $p\sim\exp\left(
-\Delta E/kT\right)$, where $\Delta E$ indicates energy difference
between initial and final configuration of a spin, on the other
hand, $k$ and $T$ then are the Boltzmann constant and temperature
values, respectively. The energy difference is then given by $\Delta
E=E_{f}-E{i}$ where $E_{f}$ and $E_{i}$ represent final and initial
energy of a spin, respectively. The standard Metropolis MC Algorithm
follows those steps: i) initially all spins are randomly set, ii) a
trial configuration is made by sequentially choosing one spin, iii)
the energy of a choosing spin is calculated due to $\Delta E$, iv)
if $\Delta E<0$ then the new state of spin is accepted (i.e.,
choosing spin flips), otherwise, v) a random number $r$ in the unit
interval is generated and the new state is only accepted if $r\leq
\exp\left( -\Delta E/kT\right)$, otherwise, the previous state is
retained, and finally, vi) the value of desired quantities are
determined. Whereas, in our simulation algorithm approach, we have
modified the Boltzmann probability function as $p_{i}\sim\exp\left(
-E_{i}/kT\right)$ where $E_{i}$ represents initial energy of a spin
$i$ defined Eq.\,(6). After initially the probability $P_{o}$ has
been set as zero which indicates that there is no up spin (i.e, open
channel) in the lattice, our method follows those steps: i) all
spins are initially set as down, i.e., all channels are closed, ii)
a trial configuration is made by sequentially choosing one spin,
iii) the energy of choosing spin is calculated due to Eq.\,(6). iv)
a random number $r$ in the unit interval is generated and the new
state is only accepted if $r\leq\exp\left( -E_{i}/kT\right)$,
otherwise, the previous state is retained, and finally, v) the value
of desired quantities are determined. As seen above steps, a
choosing spin flips only depends on probability with
$p_{i}\sim\exp\left( -E_{i}/kT\right)$. Using this modified
algorithm, simulation of the above model has been performed on an
$L\times L$ square lattice ($L=20$) with periodic boundary
conditions. In the simulation $k$ was set as unity, and the data
were also generated with $5000$ MC steps per site.

\section{Simulation Results}

In our simulation we consider the BTX-modified Na+ channels in the
squid giant axon \cite{Correa}. The probability of channel opening
($P_{o}$) for non-interactive channel systems are well described by
a Boltzmann distribution of the form as a function of voltage and
temperature \cite{Aidley},
\begin{equation}
P_{o}=\left\{  1+\exp\left[  -ze_{0}\left(  V-V_{0}\right)
/kT\right] \right\}  ^{-1}
\end{equation}
where $z$ (number of charges on gating particle) and $V_{0}$ (the
voltage at which half of the channels are open) are called the
Boltzmann parameters (or gating parameters), $k$ is the Boltzmann
constant and $T$ is the temperature of the system. The parameters
$z$ and $V_{0}$ are found from different single channel experiments.
A single Boltzmann distributions was considered to fit all the data
available from experiments done at different temperatures and in
different axons. Particularly, sigmoidal-shaped curves were obtained
for the steady-state channel opening of the BTX-modified sodium
channels by Correa et al. \cite{Correa}. They determined $P_{o}$
from experimental recordings made at the bath temperatures $0,$
$3.4,$ $3.7,$ $8.5,$ $14.3$ $^{o}C$. For each temperature,
recordings at different voltages were made after the bath
temperature had stabilized. The fits to Boltzmann distributions gave
values equal to $-67.5,$ $-63.9,$ $-66.8,$ $-61.2,$ $-57.6$ $mV$ and
$z$ values equal to $3.7,$ $3.6,$ $3.6,$ $4.0,$ $4.1$ for
temperatures $0,$ $3.4,$ $3.7,$ $8.5,$ $14.3$ $^{o}C$, respectively.
\begin{figure} \label{fig:1}
\includegraphics[width=12cm, height=10cm]{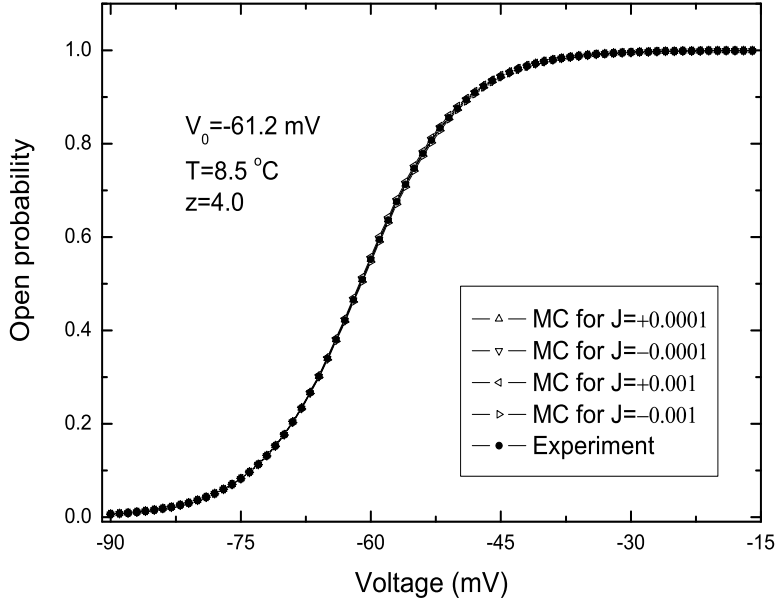}
\caption{\label{fig:epsart} Probability of channel opening ($P_{o}$)
at $V_{0}=-61.2$ $mV$, $T=8.5$ $^{o}C$, $z=4.0$, $J=\pm0.0001$ and
$J=\pm0.001$ $eV$; open triangles: MC calculations; filled circle:
experimental data Ref.\,\cite{Correa}.}
\label{fig:1}   
\end{figure}
\begin{figure} \label{fig:2}
\includegraphics[width=12cm, height=10cm]{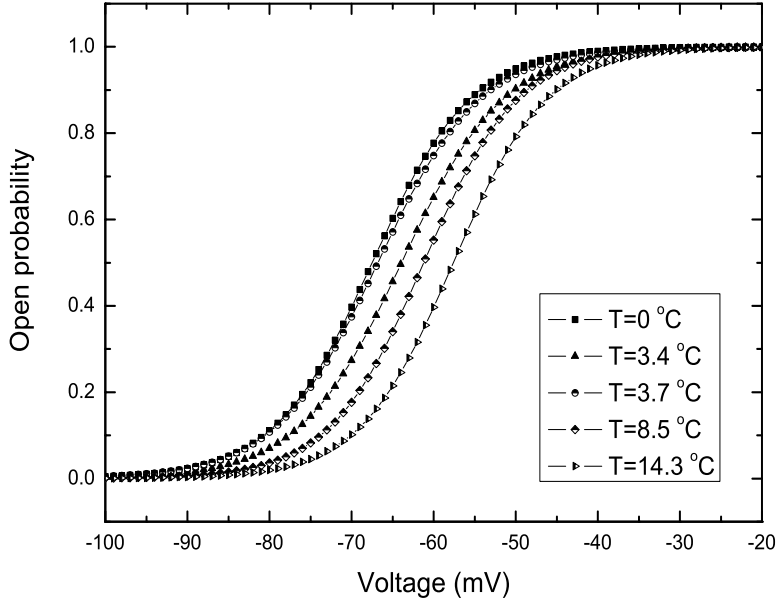}
\caption{\label{fig:epsart} Voltage dependence of the probability of
opening ($P_{o}$) of a BTX sodium channel at several temperatures
for the two-dimensional lattice with $J=0.0001$ $eV$. The curves are
obtained using the same parameters of Ref.\,\cite{Correa}.}
\end{figure}

Fig.\,1 shows the $P_{o}$ as a function of voltage for the case
$T=8.5$ $^{o}C$, $z=4.0$, $V_{0}=-61.2$ $mV$. To compare our MC
results using these parameters with the experiment we also show the
experimental data for the same parameters in Fig.\,1. In this
figure, the open-triangles and filled-circle correspond to MC
simulation calculations and experimental measurement, respectively.
In the case of weakly interacting channels (with $J=\pm0.0001$ and
$J=\pm0.001$ $eV$), a matching of the theoretically computed data
based on the MC simulation with the experimental observations is
seen explicitly. It is found that the slope value of the sigmoidal
curve at $V_{0}$ is exactly equal to its value at $V_{0}$ for the
experimental curve. These results well agree with the
two-dimensional model predictions for the ligand-gated ion channels
\cite{Liu}. In the two-dimensional model of ligand-gated ion
channels, expression for the mean open probability, $P_{o}$, was
derived from the grand partition function and interactions between
neighboring open channels gave rise to sigmoidal $P_{o}$ vs.
concentration curve.  The same Hill slope was found in the model as
the fit of the experimental data obtained from nicotine ACh receptor
channels.

The dependence of the probability of being open state ($P_{o}$) on
the membrane voltage ($V$) is shown in Fig.\,2 for a two-dimensional
lattice, using the values of  $z$, $V_{0}$, temperature ($T$) given
in Ref.\,\cite{Correa} with $J=0.0001$ $eV$. Analysis of the $P_{o}$
data shows that the probability of being open vs. voltage relation
is also sigmoidal, that is the fraction of channels in open state
rises to $1.0$ from $0.0$, as in the HH model \cite{Hodgkin}. From
the figure one can see that increasing temperature causes $P_{o}$
vs. $V$ curve to shift to the right along the voltage axis without
changing the voltage dependence. The direction of the shift implies
that an increase in temperature stabilizes the closed or
destabilizes the open configuration of the channel, because large
depolarizing voltage is needed to obtain the same $P_{o}$. These
results are in good agreement with the results reported in gating
kinetics of BTX-modified sodium channels \cite{Correa}, and confirm
that our model is correct in MC simulation framework to simulate
open probability of BTX channels.
\begin{figure} \label{fig:3}
\includegraphics[width=12cm, height=10cm]{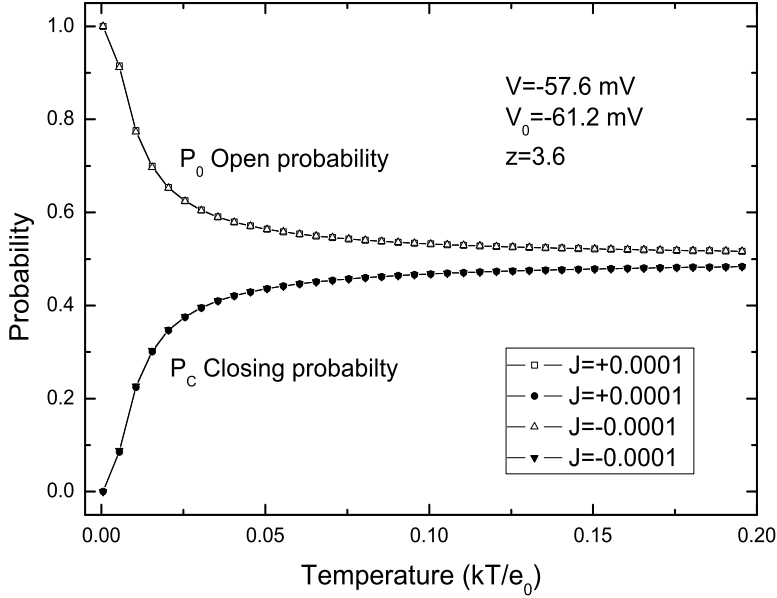}
\caption{\label{fig:epsart} Opening and closing probabilities of
channels vs. temperature $kT/e_{0}$ at $J=\pm0.0001$, $V=-57.6$
$mV$, $V_{0}=-61.2$ $mV$, $z=3.6$.}
\end{figure}
\begin{figure} \label{fig:4}
\includegraphics[width=12cm, height=10cm]{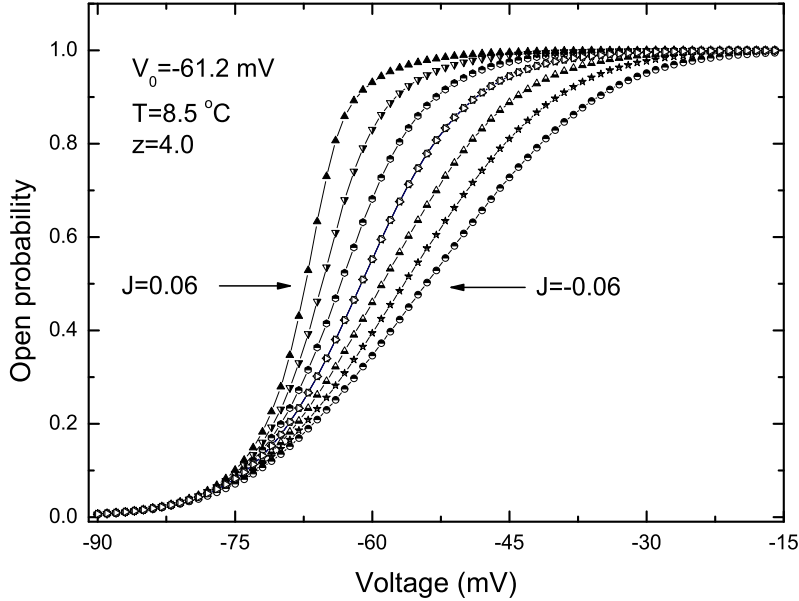}
\caption{\label{fig:epsart} MC results for different values of the
energy of interaction ($J$) between adjacent open channels at
$V_{0}=-61.2$ $mV$, $T=8.5$ $^{o}C$, $z=4.0$ for $J=\pm0.0001$,
$\pm0.02$, $\pm0.04$, $\pm0.06$ $eV$.}
\end{figure}
On the other hand, in Fig. 3, we demonstrated the simulation results
for the temperature dependence of opening and closing probabilities
of the channels. This figure represents the equilibrium behavior of
ion channels on the membrane at different temperature. As seen from
Fig.\,3 opening probability $P_{o}$ of ion channels exponentially
decreases with increasing temperature from $0.0$ to $0.2$.
Conversely, closing probability $P_{c}$ of ion channels
exponentially increases with increasing temperature. However,
functional behavior of opening and closing probabilities obey second
order exponential form as $P_{o,c}=\pm A_{1}\exp\left(\mp
T/\tau_{1}\right)\pm A_{2}\exp\left(\mp T/\tau_{2}\right)$ with
$A_{1}=0.45$, $A_{2}=0.07$ and $\tau_{1}=0.01$, $\tau_{2}=0.12$ for
$J=\pm0.0001$. At high temperatures, converging to $0.5$ of the
opening and closing probabilities indicate that this value is
equilibrium value for ion channels. That is, at high temperatures,
while the half of ion channels is approximately open, the other half
is close, randomly. This is expected behavior for ion channels.

Finally, to assess the effects of strong cooperativity on $P_{o}$
vs. $V$ variation we show the MC simulations for the dependence of
open channel probability on membrane potential for, $J=\pm0.0001$,
$\pm0.02$, $\pm0.04$, $\pm0.06$ $eV$ for fixed values of $z$,
$V_{0}$ and $T$ in Fig.\,4. Two-dimensional model predicts that both
the position and shape of the curves are different when the
interaction energy ($J$) between the channels are changed; e.g.,
when open channels exhibit positive cooperativity ($J>0$), $P_{o}$
vs. $V$ curve is shifted to lower voltages and the slope at $V_{0}$
becomes steeper. When $J<0$, the curve is shifted to higher voltages
and the slope becomes less steep, seen in Fig.\,4. Similar behaviors
have also been observed in two-dimensional model predictions for the
ligand-gated ion channels \cite{Liu}. Positive cooperativity
increased the slope at the midpoint of the $P_{o}$ vs. concentration
curve, shifted the apparent binding affinity to lower
concentrations. Negative cooperativity had the opposite effects.
Thus, the present modified Hamiltonian and modified MC simulation
algorithm may be useful approach to test the results reported in
Ref.\,\cite{Liu}.

\section{Summary and Conclusion}

In this study, we have presented the MC simulation for statistical
mechanics model of cooperativity between voltage-sensitive ion
channels in cell membranes. We have also introduced a new
Hamiltonian and a modified MC simulation algorithm for calculating
the fraction of channels in open state, i.e., open probability at
equilibrium for the channels. The new Hamiltonian contains global
coupling between the channels and the algorithm is an improvement
over the standard metropolis MC algorithm. Since the global or
long-range interactions among the sodium channels in squid giant
axons are crucial for generation of action potentials and the BTX is
able to induce profound changes in the behavior of nerve sodium
channels, we have considered the BTX-modified sodium channels of
squid giant axon in our simulation. Open probability for the
BTX-modified sodium channels have been found for different voltage,
temperature and coupling strength to other channels (J). It was
shown that in the case of weak cooperativity ($J\approx0$) the
results well match the experimental data using cut-open axon
technique. On the other hand, in the case of strong cooperativity
($J>0$), the MC simulation predicted a different sort of voltage
dependence for the open probability, i.e., a shift along the voltage
axis and a change in slope at the midpoint potential. When compared
with model curves for the ligand-gated channels the simulation
results also well agree with the two-dimensional Ising model
predictions for the open probability versus concentration curves.
This indicates that the present simulation method is applicable to
the ligand gated channel systems.

\begin{acknowledgments}
One of the authors (R.E.) thanks Professors T. Hakioglu and U.
Tirnakli for their help in his participation in the Summer School on
Nonextensive Statistical Mechanics and Complexity (July, 2007,
Institute of Theoretical and Applied Physics, Marmaris, Turkey) and
their hospitality extended to him during the first stages of this
work.
\end{acknowledgments}

\end{document}